\documentstyle[a4,11pt]{article}

\begin{document}
\hfill
NIKHEF-H/92-17

\begin{center}
{\huge The three-loop QCD $\beta$-function and anomalous dimensions} \\
\vspace{1cm}
{\bf S.A. Larin}
\footnote{on leave from Institute for Nuclear Research (INR)
of the Russian Academy of Sciences, Moscow 117312, USSR.}
{\bf and}
{\bf  J.A.M.Vermaseren} \\
\vspace{2mm}
NIKHEF-H \\ P.O. Box 41882 \\ 1009 DB Amsterdam. \\ \vspace{5mm}
\end{center}

\begin{abstract}
The analytic calculation of the three-loop QCD $\beta$-function and
anomalous dimensions within the minimal subtraction scheme
in an arbitrary covariant guage is presented.
The result for the $\beta$-function coincides with the previous
calculation \cite{tvz}.
\end{abstract}

\newpage
The calculation of the one-loop $\beta$-function in Quantum
Cromodynamics (QCD) has lead to the discovery of the asymptotic freedom
in this model and to the establishing QCD as the theory of strong interactions
\cite{gvp}.
The two-loop QCD $\beta$-function was calculated in \cite{2l}.
The first calculation of the three-loop $\beta$-function in
QCD was done in \cite{tvz}.
Since then the perturbative calculations of physical quantities in QCD
reached the $\alpha_s^3$ (next-next-to-leading) order of perturbative
expansion in the strong coupling constant $\alpha_s$
for the total cross section of $e^+e^-$ annihilation into hadrons
\cite{gkl} and for the deep inelastic lepton-hadron scattering \cite{tlv}.
The phenomenological application of these calculations involves
the three-loop approximation of the $\beta$-function. The three-loop
$\beta$-function is also necessary for the investigation of the
scheme-dependence problem in QCD and infrared
behavior of the coupling constant \cite{ckl},\cite{stev}.
That is why it is important to have an independent check of such
a complicated calculation as the calculation of the three-loop QCD
$\beta$-function \cite{tvz}.

In this paper we present original three-loop
results for QCD anomalous dimensions in an arbitrary covariant gauge
and confirm the three-loop result for the $\beta$-function of the
work \cite{tvz}.

Throughout the calculations we use the dimensional regularization \cite{hv}
and the minimal subtraction (MS) scheme \cite{h}.
The definition of the $\beta$-function is:
\begin{eqnarray}
\label{defbeta}
\beta (a) & = & \mu^2 \frac{\partial a}{\partial \mu^2}
              = - \sum_{i=0}^{\infty} \beta_ia^{i+2},
\end{eqnarray}
here $a=g^2/16\pi^2=\alpha_s/4\pi$ and $g$ is the strong
coupling constant in the standard QCD lagrangian.
To calculate the $\beta$-function we need to calculate those renormalization
constants of the lagrangian which determine the renormalization constant
of the charge.
It was convenient for us (for technical reasons) to choose the following
three renormalization constants: $Z_1$ of the quark-quark-gluon vertex
and correspondingly
$Z_2$ of the inverted quark propagator and $Z_3$ of the inverted gluon
propagator. Note that the authors of the work \cite{tvz} used another
combination: ghost-ghost-gluon vertex and correspondingly inverted
ghost and gluon propagators. So our calculation provides a really
independent check of the three-loop $\beta$-function.

Renormalization constants within MS-scheme do not
depend on dimensional parameters (masses, momenta) \cite{collins} and
have the following structure:
\begin{eqnarray}
\label{zet}
Z(a)=1+\sum_{n=1}^{\infty}\frac{z^{(n)}(a)}{\epsilon^n},
\end{eqnarray}
where $\epsilon$ is the parameter of the dimensional regularization,
the dimension of the space-time being $D=4-2\epsilon$.
The renormalization constants define corresponding anomalous dimensions:
\begin{eqnarray}
\label{defga}
\gamma(a)=\mu^2\frac{d\log Z(a)}{d\mu^2} =
-a\frac{\partial z^{(1)}(a)}{\partial a}.
\end{eqnarray}
It is convenient to express the $\beta$-function through the corresponding
anomalous dimensions:
\begin{eqnarray}
\label{betaviaga}
\beta(a) &=& a~[~\gamma_3(a)+2\gamma_2(a)-2\gamma_1(a)~].
\end{eqnarray}
The calculation of the renormalization constants within MS-scheme can
be reduced to the calculation only
of massless propagator diagrams by means of the method of
infrared rearrangement \cite{vladimirov}.
In the case of the quark-quark-gluon vertex
it means in particular that we can safely nullify the gluon momentum reducing
the calculation of the $Z_1$ to the propagator diagrams.

One way to find the constants $Z_i$ is to obtain them
as the sums of counterterms
of individual diagrams. This implies application of R-operation to each
individual diagram which is an enormous amount of work in the three-loop case.
This way was used in \cite{tvz}.
But now we have en efficient program Mincer \cite{mincer} written for
the symbolic manipulation system Form \cite{form}.
This program computes analytically three-loop massless propagator diagrams
within dimensional regularization. So we are able to obtain renormalization
constants at the three-loop level straightforwardly from the
multiplicative renormalizability of Green functions:
\begin{eqnarray}
\label{multren}
\Gamma_{Renormalized}(a)=Z(a)~\Gamma_{Bare}(a_{B}),
\end{eqnarray}
where $a_{B}$ is the bare charge.

Thus we compute with the program Mincer the unrenormalized three-loop
one-particle-irreducible quark-quark-gluon vertex
and inverted quark and gluon propagators.
Having the two-loop
bare charge we determine the necessary three-loop constants
$Z_1$, $Z_2$ and $Z_3$ from the requirement that the poles
in $\epsilon$ cancel in the r.h.s. of eq.(\ref{multren}).
The computations were done in the covariant gauge with an arbitrary gauge
parameter $\xi$ to check directly the gauge independence
of the $\beta$-function.
So the diagrams were run with the gluon propagator
$(g_{\mu\nu}-\xi\frac{q_\mu q_\nu}{q^2})/q^2$.
(Note that the computations in \cite{tvz} were done in the Feynman gauge
$\xi=0$.)

The results of our three-loop computation of the anomalous dimensions
in the MS-scheme are:
\begin{eqnarray}
\label{gamma1}
\gamma_1(a) &=& +a~[~  C_F + C_A +\xi (  - C_F - 1/4C_A )~]
\nonumber \\ &&
    +a^2~[~- \frac{3}{2}C_F^2 +\frac{17}{2}C_FC_A+ \frac{67}{24}C_A^2
- C_Fn_f  -\frac{5}{12}C_An_f
\nonumber \\ &&
       ~~~+ \xi (  - \frac{5}{2}C_FC_A - \frac{15}{16}C_A^2 )
       + \xi^2 ( \frac{1}{4}C_FC_A + \frac{1}{8}C_A^2 )~]
\nonumber \\ &&
    +a^3~[~ \frac{3}{2}C_F^3 +C_F^2C_A (12\zeta_3 - \frac{143}{4})
  +C_FC_A^2 (- \frac{15}{2}\zeta_3+ \frac{10559}{144})
  + C_A^3(\frac{3}{4}\zeta_3+ \frac{10703}{864})
\nonumber \\ &&
{}~~~ + \frac{3}{2}C_F^2n_f +C_FC_An_f (6\zeta_3- \frac{853}{36})
  +C_A^2n_f (- \frac{9}{2}\zeta_3 - \frac{205}{216})
  + \frac{5}{9}C_Fn_f^2- \frac{35}{108}C_An_f^2
\nonumber \\ &&
{}~~~ + \xi ( C_FC_A^2( - \frac{3}{2}\zeta_3- \frac{371}{32})
+C_A^3 (- \frac{9}{16}\zeta_3 - \frac{127}{32})
 + \frac{17}{8}C_FC_An_f  + \frac{1}{2}C_A^2n_f  )
\nonumber \\ &&
{}~~~ + \xi^2 (C_FC_A^2( \frac{3}{8}\zeta_3 + \frac{69}{32})
 +C_A^3( \frac{3}{32}\zeta_3+ \frac{27}{32}) )
+ \xi^3(  - \frac{5}{16}C_FC_A^2 - \frac{7}{64}C_A^3 )~],
\\
\gamma_2(a) &=& +a~[~ C_F + \xi (  - C_F )~]
\nonumber \\ &&
 +a^2~[~- \frac{3}{2}C_F^2 + \frac{17}{2}C_FC_A - C_Fn_f
       + \xi (  - \frac{5}{2}C_FC_A ) + \xi^2 ( \frac{1}{4}C_FC_A )~]
\nonumber \\ &&
 +a^3~[~ \frac{3}{2}C_F^3 +C_F^2C_A( 12\zeta_3- \frac{143}{4})
 +C_FC_A^2 (- \frac{15}{2}\zeta_3+ \frac{10559}{144})+ \frac{3}{2}C_F^2n_f
\nonumber \\ &&
{}~~~- \frac{1301}{72}C_FC_An_f  + \frac{5}{9}C_Fn_f^2
  + \xi ( C_FC_A^2( - \frac{3}{2}\zeta_3  - \frac{371}{32} )
+ \frac{17}{8}C_FC_An_f)
\nonumber \\ &&
{}~~~ + \xi^2  (C_FC_A^2( \frac{3}{8}\zeta_3 + \frac{69}{32}) )
       + \xi^3  (  - \frac{5}{16}C_FC_A^2 )~],
\\
   \gamma_3(a) &=& +a~[~ - \frac{5}{3}C_A + \frac{2}{3}n_f
       + \xi (  - \frac{1}{2}C_A )~]
\nonumber \\ &&
  + a^2~[~  - \frac{23}{4}C_A^2+2C_Fn_f + \frac{5}{2}C_An_f
       + \xi (  - \frac{15}{8}C_A^2 )
       + \xi^2 ( \frac{1}{4}C_A^2 )~]
\nonumber \\ &&
  + a^3~[ C_A^3 ( \frac{3}{2}\zeta_3- \frac{4051}{144})- C_F^2n_f
+C_FC_An_f(12\zeta_3 + \frac{5}{36})
+C_A^2n_f (- 9\zeta_3 + \frac{875}{36})
\nonumber \\ &&
{}~~~ - \frac{11}{9}C_Fn_f^2 - \frac{19}{9}C_An_f^2
\nonumber \\ &&
{}~~~ + \xi( C_A^3 (- \frac{9}{8}\zeta_3- \frac{127}{16}) + C_A^2n_f  )
       + \xi^2(C_A^3( \frac{3}{16}\zeta_3 + \frac{27}{16}) )
       + \xi^3(  - \frac{7}{32}C_A^3 )~].
\end{eqnarray}

Here $C_F=(n_c^2-1)/2n_c$ and $C_A=n_c$ are
the Casimir operators of the defining and adjoint
representations of the color group $SU(n_c)$, $n_f$ is the number of active
quark flavors, $\zeta_3$ is the Riemann zeta-function ($\zeta_3 =
1.202056903\ldots$), $\xi$ is the gauge parameter.

Substituting the obtained $\gamma$-functions to the eq.(\ref{betaviaga})
we find the desired $\beta$-function:
\begin{eqnarray}
\label{bet}
   \beta(a) &=&
       - a^2  [  + \frac{11}{3}C_A - \frac{2}{3}n_f ]
\nonumber \\ &&
       - a^3  [ \frac{34}{3}C_A^2 - 2C_Fn_f - \frac{10}{3}C_An_f  ]
\nonumber \\ &&
       - a^4  [ \frac{2857}{54}C_A^3+ C_F^2n_f - \frac{205}{18}C_FC_An_f
- \frac{1415}{54}C_A^2n_f + \frac{11}{9}C_Fn_f^2  + \frac{79}{54}C_An_f^2 ].
\end{eqnarray}
This result coincides with the previous calculation \cite{tvz}.

The $\beta$-function is gauge independent as it should be
within the MS-scheme and does not contain $\zeta_3$. The cancellation of the
$\xi$-terms is a strong check of the validity of the calculations.

We repeat the result once more for the particular case of QCD ($n_c=3$):
\begin{eqnarray}
\label{betanum}
 \beta(a) = -a^2 (   11 - \frac{2}{3}n_f )
       - a^3 (   102 - \frac{38}{3}n_f )
       - a^4 (  \frac{2857}{2} - \frac{5033}{18}n_f + \frac{325}{54}n_f^2 ).
\end{eqnarray}

It is interesting to note that the poles of individual three-loop diagrams of
the quark-quark-gluon vertex and the the total renormalized three-loop
quark-quark-gluon
vertex (which we have also calculated but do not present here because
of the cumbersome result) contain the quartic Casimir operator
(see e.g. \cite{cvit}). But it cancels in the
anomalous dimension $\gamma_1$ and correspondingly
in the $\beta$-function which
are expressed through the quadratic Casimir operators $C_F$ and $C_A$ only.
Since the quartic Casimir operator is present in the finite
part of the three-loop
quark-quark-gluon vertex it can enter the four-loop $\beta$-function.
But although existing techniques \cite{vladimirov}, \cite{mincer}
are sufficient
to compute the four-loop $\beta$-function analytically,
in practice it is a very difficult task because of enormous amount of
diagrams to be calculated.

\end{document}